\documentclass[runningheads]{llncs}

\usepackage{cite}
\usepackage{amsmath,amssymb,amsfonts}
\usepackage{algorithmic}
\usepackage{graphicx}
\usepackage{textcomp}
\usepackage{xcolor}
\usepackage{threeparttablex}
\usepackage{subcaption}
\usepackage{verbatim}
\usepackage{float}
\usepackage{booktabs}
\usepackage[utf8]{inputenc}

\usepackage{hyperref}
\makeatletter
\def\UrlAlphabet{%
      \do\a\do\b\do\c\do\d\do\e\do\f\do\g\do\h\do\i\do\j%
      \do\k\do\l\do\m\do\n\do\o\do\p\do\q\do\r\do\s\do\t%
      \do\u\do\v\do\w\do\x\do\y\do\z\do\A\do\B\do\C\do\D%
      \do\E\do\F\do\G\do\H\do\I\do\J\do\K\do\L\do\M\do\N%
      \do\O\do\P\do\Q\do\R\do\S\do\T\do\U\do\V\do\W\do\X%
      \do\Y\do\Z}
\def\UrlDigits{\do\1\do\2\do\3\do\4\do\5\do\6\do\7\do\8\do\9\do\0}
\g@addto@macro{\UrlBreaks}{\UrlOrds}
\g@addto@macro{\UrlBreaks}{\UrlAlphabet}
\g@addto@macro{\UrlBreaks}{\UrlDigits}
\makeatother

\NewEnviron{myequation}{%
    \begin{equation}
    \scalebox{0.9}{$\BODY$}
    \end{equation}
    }
    
\usepackage{array}
\renewcommand{\arraystretch}{1.2}
\newcolumntype{?}{!{\vrule width 0.7pt}}

\usepackage{multirow}
\usepackage{multicol}

\usepackage[font=small,labelfont=bf,tableposition=top]{caption}
\DeclareCaptionLabelFormat{andtable}{#1~#2  \&  \tablename~\thetable}

\def\BibTeX{{\rm B\kern-.05em{\sc i\kern-.025em b}\kern-.08em
    T\kern-.1667em\lower.7ex\hbox{E}\kern-.125emX}}
\begin{document}

\title{HPC AI500: Representative, Repeatable and Simple HPC AI Benchmarking\\
}

\author{Zihan Jiang\inst{1,2},  Wanling Gao\inst{1,2,3}, Fei Tang\inst{1,2}, Xingwang Xiong\inst{1,2}, Lei Wang\inst{1,2,3},  Chuanxin Lan\inst{1}, Chunjie Luo\inst{1}, Hongxiao Li\inst{1,2}, Jianfeng Zhan\inst{1,2,3}}
\authorrunning{ZH. Jiang et al.}
\institute{Institute of Computing Technology, Chinese Academy of Sciences \and
University of Chinese Academy of Science \\
\and International Open Benchmark Council (BenchCouncil)\\
\email{\{jiangzihan, gaowanling, tangfei, wanglei\_2011, xiongxingwang, lanchuanxin, luochunjie, lihongxiao, zhanjianfeng\}@ict.ac.cn}
}
\maketitle

\begingroup\renewcommand\thefootnote{\textsection}
\footnotetext{Jianfeng Zhan is the corresponding author.}
\endgroup

\begin{abstract}

Recent years witness a trend of applying large-scale distributed deep learning algorithms (HPC AI) in both business and scientific computing areas, whose goal is to speed up the training time to achieve a state-of-the-art quality. 
The HPC AI benchmarks accelerate the process. Unfortunately, benchmarking HPC AI systems at scale raises serious challenges.
This paper presents a representative, repeatable and simple HPC AI benchmarking methodology. Among the seventeen AI workloads of AIBench Training---by far the most comprehensive AI Training benchmarks suite---we choose two representative and repeatable AI workloads. The selected HPC AI benchmarks include both business and scientific computing: Image Classification and Extreme Weather Analytics. To rank HPC AI systems, we present a new metric named Valid FLOPS, emphasizing both throughput performance and a target quality.
The specification, source code, datasets, and HPC AI500 ranking numbers are publicly available from \url{https://www.benchcouncil.org/HPCAI500/}.

\end{abstract}

\keywords{
HPC AI, Distributed Deep Learning, Benchmarking, Metric}

\section{Introduction}

\captionsetup[figure]{font=scriptsize}
\captionsetup[table]{font=scriptsize}

The massive success of AlexNet~\cite{krizhevsky2017imagenet} in the ImageNet~\cite{deng2009imagenet} competition marks the booming success of deep learning (DL) in a wide range of commercial application areas.
Like image recognition and natural language processing, many commercial fields achieve unprecedented accuracy, even outperforming ordinary human beings' capability. Though it is much more challenging to obtain high-quality labeled scientific data sets, there is an increasing trend in applying DL in scientific computing areas~\cite{kurth2018exascale,racah2017extremeweather, Jia2020Pushing}.

With massive training data available, recent years witness a trend of applying distributed DL algorithms at scale in commercial and scientific computing areas. Motivated by these emerging HPC AI workloads, the HPC community feels interested in building HPC AI systems to reduce time-to-quality, which depicts the training time to achieve a target quality (e.g., accuracy). For example, several state-of-the-practice HPC AI systems ~\cite{kurth2018exascale, Jia2020Pushing} are built to tackle enormous AI challenges. 
The benchmark accelerates the process~\cite{hennessy2011computer,mattson2020mlperf}, as it provides not only design inputs but also evaluation and optimization metrics and methodology~\cite{tang_ISPASS_2021, gao2018aibench,mattson2020mlperf}. However, there are several challenges in benchmarking HPC AI systems.

The first challenge is how to achieve both representative and simple, which essential properties the past successful benchmark practices establish.  
On the one hand, the SPECCPU~\cite{speccpu}, PARSEC~\cite{bienia2008parsec}, and TPC benchmarks, like TPC-DS~\cite{tpcds} emphasize the paramount importance~\cite{tang_ISPASS_2021} of the benchmarks' being representative, and diverse, as no single benchmark or metric measures the performance of computer systems on all applications~\cite{gray1993database}.

On the other hand,  TOP500~\cite{dongarra1997top500} establishes the de facto supercomputer benchmark standard in terms of simplicity. Simplicity has three implications: first, the benchmark is easy to port to a new system or architecture; second, the benchmarking cost is affordable for measuring systems at scale; third, the number of the metric is not only linear, orthogonal, and monotony~\cite{dongarra1997top500}, but also easily interpretable and understandable.



In the AI domain, there are massive AI tasks and models with different performance metrics. For example, by far, the most comprehensive and representative AI benchmark suite--AIBench~\cite{tang_ISPASS_2021, gao2018aibench} contains seventeen AI tasks.  It is not affordable to implement so many massive benchmarks and further perform benchmarking at scale. What criteria are for deciding the representative and simple benchmarks that can measure the HPC AI systems fairly and objectively? 

Second, the benchmark mandates being repeatable, while AI's nature is stochastic, allowing
multiple different but equally valid solutions~\cite{mattson2020mlperf}. Previous work manifests HPC AI's uncertainty by run-to-run variation in terms of epochs-to-quality and the effect of scaling training on time-to-quality~\cite{mattson2020mlperf}.

None of the previous HPC AI benchmarks simultaneously achieve representative, repeatable, and simple.
They either are not representative~\cite{deep500} or even irrelevant to HPC AI workloads in terms of kernel functions~\cite{hplai}, or overlook the differences of HPC AI workloads between scientific and business computing~\cite{mattson2020mlperf}.
This paper presents HPC AI500--a comprehensive HPC AI benchmarking methodology, tools, 
and metrics. Compared to our previous position paper~\cite{jiang2018hpc}, this paper proposes a  brand-new benchmarking methodology that simultaneously achieves representative, repeatable and simple.

We quantify the characteristics of AI models and micro-architecture and perform further randomness analysis. Among seventeen AI workloads of AIBench Training, we choose two representative and repeatable benchmarks: Image Classification (business computing) and Extreme Weather Analytics (EWA), to measure HPC AI systems. Image Classification and  EWA achieve state-of-the-art quality on the ImageNet dataset (business computing) and the  EWA dataset (scientific computing). These two benchmarks represent two clusters of AI benchmarks of AIBench Training from perspectives of computing areas, 
diversity of model complexity,
the computational cost 
and micro-architecture characteristics.
To rank HPC AI systems, 
we propose valid FLOPS, 
emphasizing the vital importance of achieving state-of-the-art quality and an additional metric--time-to-quality.
Our benchmarks simultaneously achieve representative, repeatable and simple.

\begin{figure}[ht]
\begin{subfigure}{.44\textwidth}
  \centering
  \includegraphics[width=.8\linewidth,height=0.5\linewidth]{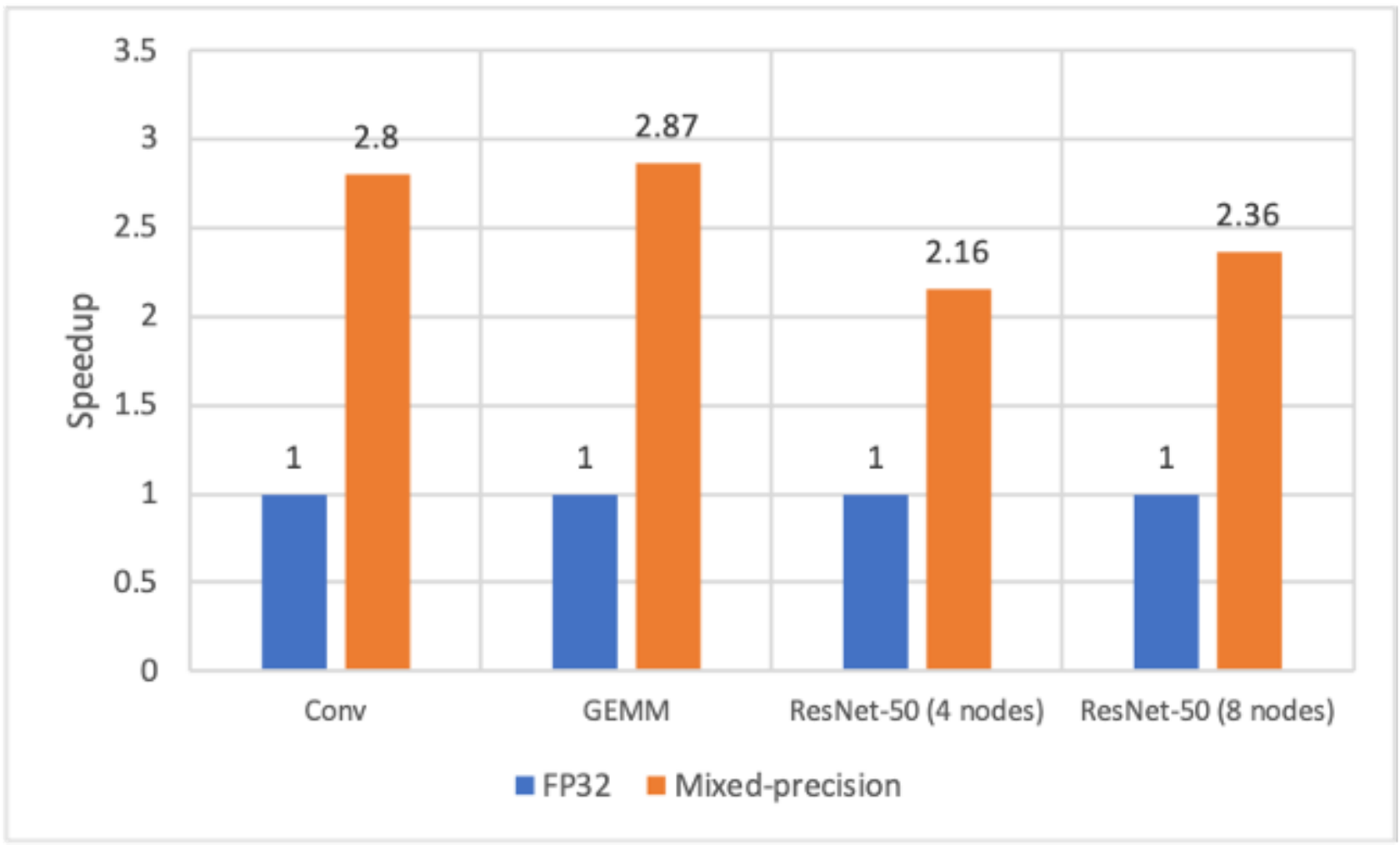}  
  \label{fig:mixed-precision-improvement}
\end{subfigure}
\begin{subfigure}{.44\textwidth}
  \centering
  \includegraphics[width=.95\linewidth,height=0.55\linewidth]{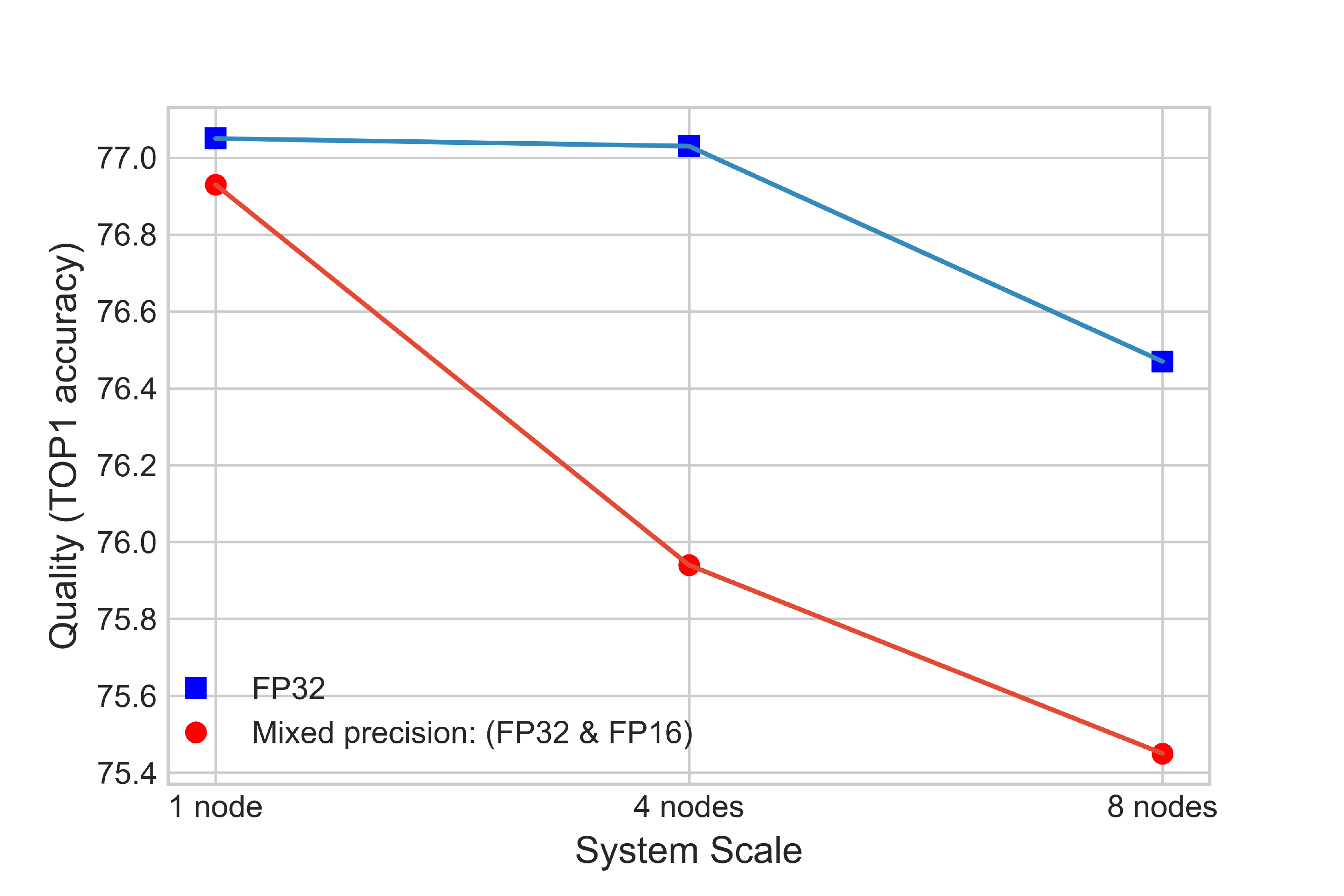}  
  \label{fig:mixed-accuracy-drop}
\end{subfigure}
\caption{Against the FP32 implementation, the mixed-precision version speeds up more than 2x the FLOPS of two micro benchmarks (Convolution and GEMM) and ResNet-50. Still, it incurs ResNet-50's accuracy loss when the system scale increases:  0.12\% at one node while about 1\% at both 4 and 8 nodes.}

\label{fig:micro_mixed_optimization}
\vspace{-0.8cm}
\end{figure}

\begin{figure}[ht]
  \centering
  \includegraphics[width=.8\linewidth]{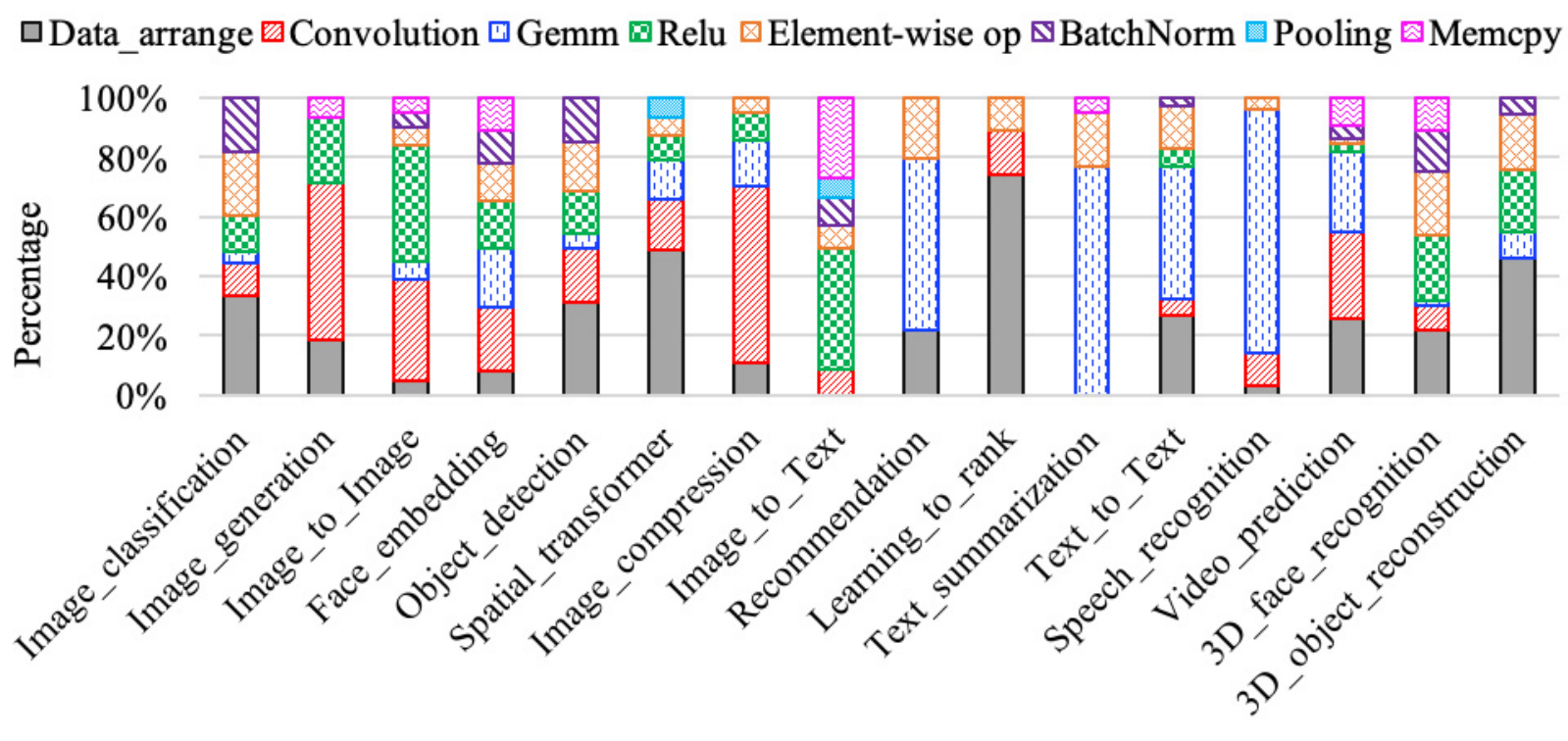}
  \caption{The kernel function breakdown of the 17 representative AI workloads from AIBench Training~\cite{tang_ISPASS_2021}; indicates the LU factorization is a not representative kernel. }
  \label{fig:breakdown_AIBench}
  \vspace{-0.5cm}
\end{figure}

\section{Motivation}~\label{challenge}

TOP500~\cite{dongarra1997top500} defines two distinctive characteristics of the de facto supercomputer benchmark standard:  simple and scalable.  We have discussed the implications of simplicity in the previous section.  Scalable means the benchmarks can measure the systems with different scales.



In the AI domain, there are massive AI tasks and models with different performance metrics. For example,  AIBench training~\cite{tang_ISPASS_2021}
contains seventeen representative AI tasks, covering a diversity of AI problem domains. 
It is not affordable for HPC AI benchmarking to implement so many massive benchmarks and further perform benchmarking at scale. 
The traditional micro or kernel benchmarking methodology,  widely used in the HPC community, can lead to the misleading conclusion, as the mixed-precision optimizations indeed improve the FLOPS of a micro benchmark like convolution while significantly impact time-to-quality
of an AI task like Image Classification. Fig.~\ref{fig:micro_mixed_optimization} shows that the mixed-precision implementation increases the FLOPS of both micro and component benchmarks but incurring an accuracy loss as the system scale increases.

The representativeness of a benchmark indicates that it must measure the peak performance and price/performance of systems when performing typical operations within that problem domain~\cite{gray1993benchmark}.
The micro benchmark, like HPL-AI~\cite{hplai}, which only contains LU decomposition, is affordable to perform a fair comparison of competing systems by isolating hardware and software from statistical  optimizations~\cite{mattson2020mlperf}. However, we found it can not represent most of the AI workloads in AIBench. As shown in Fig. ~\ref{fig:breakdown_AIBench}, the dominated kernel functions are convolution and matrix multiplication.

\section{Benchmarking Methodology}

This section introduces our benchmarking methodology. We firstly conduct a series of experiments to prove why our methodology can guarantee representativeness, simplicity, and repeatability (Sec.~\ref{subsec-achieve-repre} and Sec.~\ref{subsec-achieve-repeat}). Then we finalize the HPC AI500 benchmark decision considering these analyses and the additional requirements in the HPC field (Sec.~\ref{subsec-final-decision}).



\subsection{How to Achieve Representative}\label{subsec-achieve-repre}

We choose AIBench Training~\cite{tang_ISPASS_2021, gao2018aibench}---the most comprehensive AI benchmark by far---as the starting point for the design and implementation of HPC AI benchmarks. The experimental results of AIBench Training~\cite{tang_ISPASS_2021} have demonstrated that the seventeen AI tasks are diverse in terms of model complexity, computational cost, convergent rate, and microarchitecture characteristics covering most typical AI scenarios. To achieve representative, we identify the most typical workload in AIBench Training from both microarchitecture-independent and microarchitecture-dependent perspectives.

From the microarchitecture-dependent perspective, we choose five micro-architectural metrics to profile the computation and memory access patterns of AIBench Training, including achieved occupancy, ipc\_efficiency, gld\_efficiency, gst\_efficiency, and dram\_utilization~\cite{nvprof}. GPU architecture contains multiple streaming multiprocessors (SM); each SM has a certain number of CUDA cores, registers, caches, warp schedulers, etc. Achieved\_occupancy represents the ratio of the average active warps per active cycle to the maximum number of warps provided by a multiprocessor. Ipc\_efficiency indicates the ratio of the executed instructions per cycle to the theoretical number. Gld\_efficiency and gst\_efficiency represent the ratio of requested global memory load/store throughput to required global memory load/store throughput, respectively. 

\begin{figure}[tb]
     \centering
     \begin{subfigure}[b]{0.8\textwidth}
         \centering
         \includegraphics[scale=0.45]{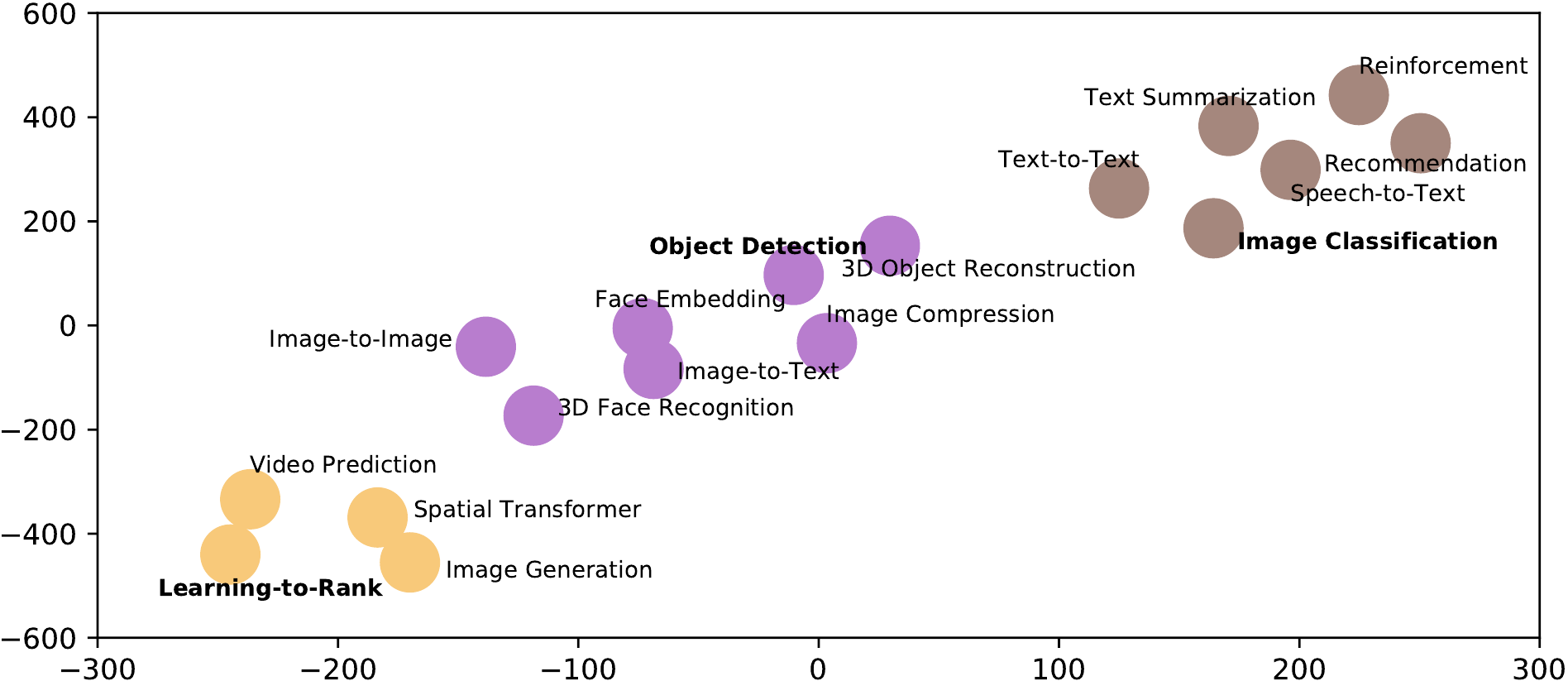}
         \caption{The microarchitecture-dependent of AIBench Training.}
         \label{fig:dependent_clustering_seventeen_component_benchmarks}
     \end{subfigure}
     \hfill
     \begin{subfigure}[b]{0.8\textwidth}
         \centering
         \includegraphics[scale=0.45]{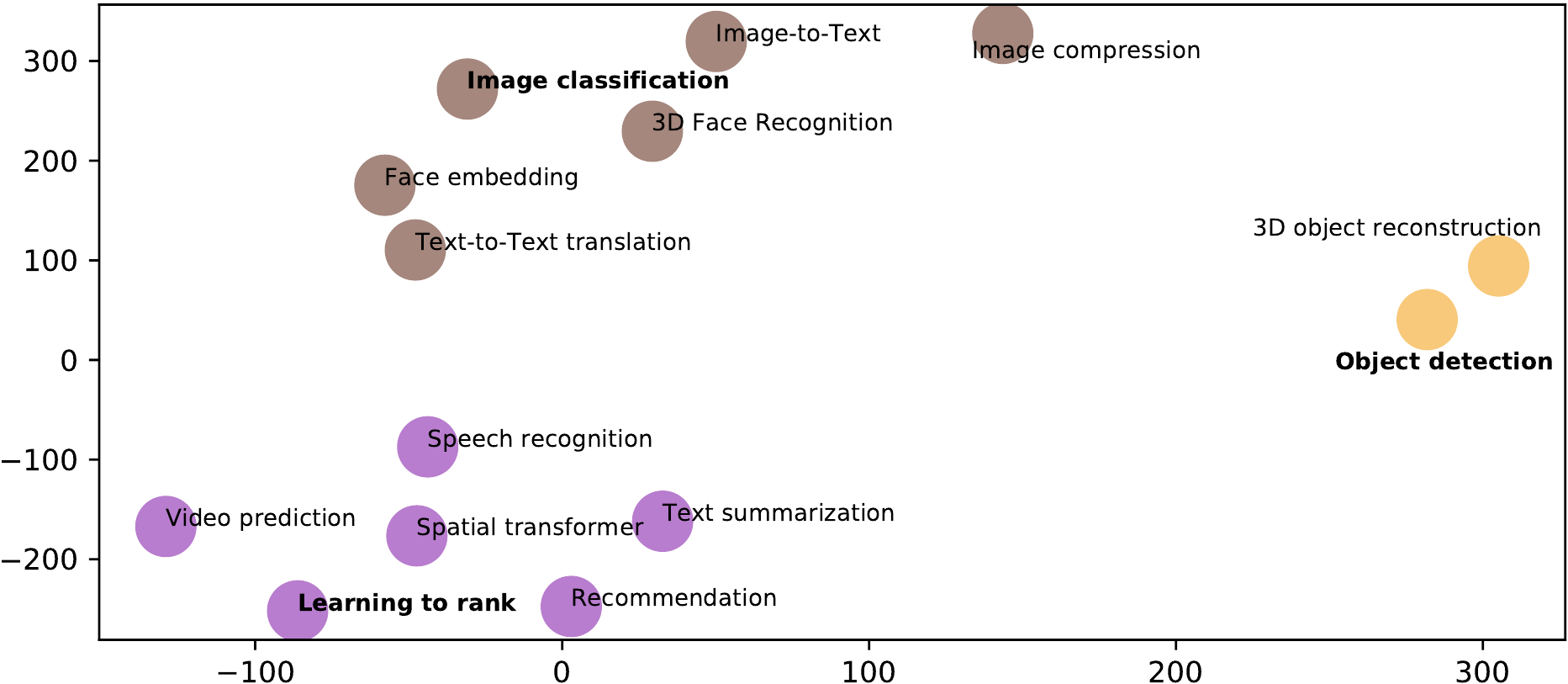}
         \caption{The microarchitecture-independent (Fig.~\ref{fig:independent_clustering_seventeen_component_benchmarks}) Clustering of AIBench Training. Image-to-image, Image Generation are not clustered due to the lack of widely accepted metrics to determine the end of a training session. NAS~\cite{zoph2016neural} probabilistically searches the network structure in each training session, resulting in unstable computational cost (FLOPs). Therefore, NAS is not included in the clustering result.} 
         \label{fig:independent_clustering_seventeen_component_benchmarks}
     \end{subfigure}
        \caption{The microarchitecture-dependent (Fig.~\ref{fig:dependent_clustering_seventeen_component_benchmarks}) and microarchitecture-independent (Fig.~\ref{fig:independent_clustering_seventeen_component_benchmarks}) Clustering of AIBench Training. The x-axis and y-axis are the positions of the Euclidean space after using t-SNE technique for visualization.}
        \label{fig:clustering-AIBench}
\end{figure}

We profile the above five metrics and perform K-means clustering on all seventeen benchmarks to explore their similarities through our TITAN XP GPUs' experiments. Note that the operating system is Ubuntu 16.04 with the Linux kernel 4.4, and the other software versions are PyTorch 1.10, python 3.7, and CUDA 10. We further use the T-SNE~\cite{rogovschi2017t} for visualization, a dimension reduction technique to embed high-dimensional data in a low-dimensional space for visualization. 
Fig.~\ref{fig:dependent_clustering_seventeen_component_benchmarks} shows the result. The x-axis and y-axis are the Euclidean space's position after using t-SNE to process the above five metrics. We find that these seventeen benchmarks are clustered into three classes.

From the microarchitecture-independent perspective, we analyze the algorithm behaviors, including model complexity (parameter size) and convergent rate (epochs to achieve the state-of-the-art quality), and system-level behaviors, including computational cost (FLOPs), for all seventeen workloads in AIBench Training. Further, we conduct a clustering analysis using these microarchitecture-independent performance data as input. Fig.~\ref{fig:independent_clustering_seventeen_component_benchmarks} shows the clustering result. 

Combing Fig.~\ref{fig:dependent_clustering_seventeen_component_benchmarks} and Fig.~\ref{fig:independent_clustering_seventeen_component_benchmarks}, we conclude that the AIBench Traning workloads consistently cluster into three classes using  both microarchitecture-dependent and microarchitecture-independent approaches.

\subsection{How to Guarantee Repeatability}\label{subsec-achieve-repeat}

Repeatability~\cite{Repli_reproducibility} refers to the variation in repeat measurements (different runs instead of different epochs using the same benchmark implementation under the identical configurations) made on the same system under test.
A good benchmark must be repeatable. Thus, repeatability is another critical criterion to select workloads for the HPC AI500 benchmarks.  However, AI's nature is stochastic~\cite{mattson2020mlperf} due to the random seed, random data traversal, non-commutative nature of floating-point addition, etc. It is hard to avoid. Thus, most AI benchmarks exhibit run-to-run variation, even using the same benchmark implementation on the same system. Therefore, we need to ensure repeatability by choosing relatively stable workloads in various AI tasks. We perform repeatability analysis using all workloads of AIBench Training on TITAN RTX GPUs. The other experiment environment is the same with Section~\ref{subsec-achieve-repre}.

To eliminate the influence of randomness as much as possible, we fix the hyperparameters for each benchmark, i.e., batch size, learning rate, optimizer, weight decays, and repeat at least four times (maximally ten times) for each benchmark to measure the run-to-run variation. Note that our evaluation uses the random seed and does not fix the initial seed except Speech Recognition. We use the coefficient of variation--the standard deviation ratio to the mean--of the training epochs to achieve a target quality to represent the run-to-run variation.

Table.~\ref{AIBench_randomness_cost} presents the run-to-run variation of seventeen workloads of AIBench. As we see, each AI benchmark varies wildly in terms of the run-to-run variation. According to Table.~\ref{AIBench_randomness_cost}, the most random workloads are Video Prediction, Text Summarization, and Image-to-Text, and their variations reach 38.46\%, 24.72\%, and 23.52\%, respectively. For Speech Recognition, even sharing the same initial seed, the run-to-run variation still gets 12.08\%. In contrast, Object Detection, Image Classification, and Learning-to-Rank the three most repeatable workloads, and the variation is 0\%, 1.12\%, and 1.9\%, respectively. 

In Section~\ref{subsec-achieve-repre}, these three workloads (The highlighted one) are consistently classified into 
three classes using the microarchitecture-dependent approach in Fig.~\ref{fig:dependent_clustering_seventeen_component_benchmarks} and the microarchitecture-independent approach in  Fig.~\ref{fig:independent_clustering_seventeen_component_benchmarks}. Overall, \emph{Image Classification}, \emph{Learning-to-Rank}, and \emph{Object Detection} achieve both representativeness and repeatability.


\subsection{Keep the Benchmarks Simple}

Simplicity is another important criterion for benchmarking. However, benchmarking an entire training session of all seventeen workloads in AIBench Training is extremely expensive, which reaches up to 10 days according to Tang et al. ~\cite{tang_ISPASS_2021}. We emphasize that \emph{Image Classification}, \emph{object Detection}, and \emph{Learning-to-Rank} achieve not only representativeness and repeatability, but also simplicity.

\begin{table}[!htbp]
\scriptsize
\caption{Run-to-run Variation of Seventeen Benchmarks of AIBench Training. Note that Image-to-image and image generation variations are not reported due to a lack of a widely accepted metric to terminate an entire training session.}
\renewcommand\arraystretch{1.2}
\label{AIBench_randomness_cost}
\center 
\begin{tabular}{>{\centering}p{0.15\textwidth}|>{\centering}p{0.35\textwidth}>{\centering}p{0.15\textwidth}>{\centering\arraybackslash}p{0.15\textwidth}}
\hline
\textbf{No.} & \textbf{Component Benchmark} & \textbf{Variation} & \textbf{Runs} \\
\hline
DC-AI-C1 & Image classification & 1.12\% & 5 \\
\hline
DC-AI-C2 & Image generation & Not available & N/A \\
\hline
DC-AI-C3 & Text-to-Text translation & 9.38\% & 6 \\
\hline
DC-AI-C4 & Image-to-Text & 23.53\% & 5 \\
\hline
DC-AI-C5 & Image-to-Image & Not available & N/A \\
\hline
DC-AI-C6 & Speech recognition & 12.08\% & 4 \\
\hline
DC-AI-C7 & Face embedding & 5.73\% & 8 \\
\hline
DC-AI-C8 & 3D Face Recognition & 38.46\% & 4 \\
\hline
DC-AI-C9 & Object detection & 0 & 10 \\
\hline
DC-AI-C10 & Recommendation & 9.95\% & 5 \\
\hline
DC-AI-C11 & Video prediction & 11.83\% & 4 \\
\hline
DC-AI-C12 & Image compression & 22.49\% & 4 \\
\hline
DC-AI-C13 & 3D object reconstruction & 16.07\% & 4 \\
\hline
DC-AI-C14 & Text summarization & 24.72\% & 5 \\
\hline
DC-AI-C15 & Spatial transformer & 7.29\% & 4 \\
\hline
DC-AI-C16 & Learning to rank & 1.90\% & 4 \\
\hline
DC-AI-C17 & Neural Architecture Search & 6.15\% & 6 \\
\hline
\end{tabular}
\vspace{-0.5cm}
\end{table}

\subsection{The Requirements in HPC Field}

\subsubsection{Dataset}
Against other domain AI benchmarks, there are two unique differences in HPC AI benchmarking. First, the challenges of HPC AI benchmarking inherit from the complexity of benchmarking scalable hardware and software systems at scale, i.e., tens of thousands of nodes, significantly different from that of IoT~\cite{luo2020comparison} or datacenter~\cite{gao2020aibench}. On this point, we need to make the benchmark as simple as possible, which we have discussed in detail before. Second, HPC AI domains cover both commercial and high-performance scientific computing. Currently, business applications are pervasive. Because of the difficulty of recruiting qualified scientists to label scientific data, 
AI for science applications lag but is promising. In general, the scientific data are often more complicated than that of the MINST or ImageNet data: the shape of scientific data can be 2D images or higher-dimension structures with hundreds of channels, while the popular commercial image data like ImageNet often consist of only RGB~\cite{jiang2018hpc}. So \emph{we should include the scientific data} in the HPC AI benchmarks. 

\subsubsection{Computation complexity} A benchmark with a small amount of computation cannot fully utilize the performance of the HPC AI system. Therefore, \emph{we exclude Learn to Ranking} because it has the lowest computation complexity in terms of FLOPS, which is only 0.08 MFLOPs in terms of a single forward computation. According  to~\cite{tang_ISPASS_2021}, Image Classification and Object Detection are more complicated than that by one or two orders of magnitude, respectively.

\subsection{The Finalized Benchmarks Decision}\label{subsec-final-decision}
Based on the existing analysis, we can conclude that \emph{Image Classification} and \emph{Object Detection} are the final candidates to construct the HPC AI500 benchmark. We investigate the broad applications of Image Classification and Object Detection in both HPC~\cite{kurth2018exascale,kurth2017deep,mathuriya2018cosmoflow,racah2017extremeweather} and commercial field~\cite{tanakaimagenet,jia2018highly,ying2018image,akiba2017extremely}. We choose the most representative workloads and data sets from these two fields. The details about the dataset and adopted model are introduced in Sec.~\ref{sec:new-design-implementation}.

\textbf{EWA} is one of the pioneering works that uses a deep learning algorithm to replace the rules predefined by a human expert and achieve excellent results~\cite{racah2017extremeweather}. Most importantly, EWA's goal is to identify various extreme weather patterns (e.g., tropical depression), essentially ~\emph{object detection}. In 2018, a deep learning-based EWA implementation~\cite{kurth2018exascale} won the Gordon Bell Prize, which is the first AI application to win this award. 

\textbf{Image Classification} is widely used in many applications of \emph{commercial fields}, which is a fundamental task in AI research. With the development of large-scale deep learning, Image Classification has become a well-known showcase optimizing HPC AI systems~\cite{tanakaimagenet,jia2018highly,akiba2017extremely}. 




\section{Benchmark Design and Implementation}\label{sec:new-design-implementation}
In this section, we first introduce the details of the HPC AI500 benchmarks, including the model, dataset, target quality (Sec.~\ref{subsec-data-model-quality}), reference implementation (Sec.~\ref{subsec-reference-implementation}), and proposed VFLOPS metric (Sec.~\ref{sec:metrics}). 

\subsection{Models, Datasets, and Target Quality}\label{subsec-data-model-quality}
\subsubsection{EWA}

\textbf{Dataset.} The EWA dataset~\cite{racah2017extremeweather} is made up of 26-year climate data. The data of every year is available as one HDF5 file. Each HDF5 file contains two data sets: images and boxes. The images data set has 1460 example images (4 per day, 365 days per year) with 16 channels. Each channel is 768 * 1152, corresponding to one measurement per 25 square km on earth. The box dataset records the coordinates of the four kinds of extreme weather events in the corresponding images: tropical depression, tropical cyclone, extratropical cyclone, and the atmospheric river.

\textbf{Model.} Faster-RCNN targets real-time Object Detection~\cite{ren2015faster}. As one of the latest models of an RCNN family~\cite{girshick2015fast,girshick2014rich}, it deprecates the selective search used in the previous RCNN version. Instead, Faster-RCNN proposes a dedicated convolutional neural network, named region proposal network (RPN), to achieve nearly cost-free region proposals. With such a design, Object Detection is much faster. As a result, Faster-RCNN wins the 1st-place entries in ILSVRC'15 (ImageNet Large Scale Visual Recognition Competition). 

\textbf{Target Quality.}
The target quality is $MAP@[IoU=0.5]=0.35$, which is our best training result. MAP means the average precision, which is a dedicated metric for object detection. The IoU means the intersection over union to measure how much our predicted boundary overlaps with the ground truth. 

\subsubsection{Image Classification}

\textbf{Dataset.} ImageNet~\cite{deng2009imagenet} is large visual database designed for use in visual object recognition research. More than 14 million images have been hand-annotated according to the WordNet hierarchy. Both the original images and bounding boxes are provided. The data size is more than 100 GB.

\textbf{Model.} ResNet is a milestone in Image Classification~\cite{he2016deep}, marking the ability of AI to identify images beyond humans in a particular domain. The spirit of ResNet is its success in reducing the negative impact of the degradation problem.  The degradation problem means in the very deep neural network; the gradient will gradually disappear in the process of back-propagation, leading to poor performance. Therefore, with ResNet, it is possible to build a deeper convolution neural network and archive the higher accuracy. Researchers successfully build a ResNet with 152 layers. This ultra-deep model won all the awards in ILSVRC'15. 

\textbf{Target Quality.}
The target quality is $Top1$ $Accuracy = 0.763$, which is the highest accuracy by far in training ImageNet/ResNet50 at scale~\cite{ying2018image}. The Top-1 accuracy refers to that only the output with the highest probability is the correct answer.

\begin{table}[ht]
\scriptsize
\centering
\caption{The Datasets Summary of HPC AI500 Benchmarks}
\begin{tabular}[ht]{>{\centering}p{0.25\textwidth}|>{\centering}p{0.2\textwidth}>{\centering}p{0.2\textwidth}>{\centering\arraybackslash}p{0.2\textwidth}}
\toprule
\textbf{Dataset}& \textbf{Channels} & \textbf{Resolution} & \textbf{Size}\\
\midrule
The Extreme Weather Dataset~\cite{racah2017extremeweather}& 16 & 768*1052  & 558 GB   \\ \midrule
ImageNet Dataset~\cite{deng2009imagenet} & 3 & 256*256  & 137 GB  \\ 

\bottomrule
\end{tabular}
\label{table:hpcaibench_dataset}
\end{table}%



\subsection{Reference Implementation}\label{subsec-reference-implementation}
The reference implementation of HPC AI500 benchmark is summarized as shown in Table~\ref{table:benchmark_suite}. At present, we provide the implementations using TensorFlow~\cite{abadi2016tensorflow}, which is a popular deep learning framework in the HPC community. For communication, we adopt Horovod~\cite{sergeev2018horovod}
instead of the default GRPC protocol in TensorFlow, which is not extendable for large-scale cluster~\cite{mathuriya2017scaling} due to the limitation of the master-slave architecture and socket-based communication. Horovod is a library originally designed for scalable distributed deep learning using TensorFlow. It implements \textit{all\_reduce} operations using ring-based algorithms~\cite{ring-based} and other high efficient communication algorithms, widely used in the traditional HPC community.

\begin{table}[ht]
\scriptsize
\centering
\caption{HPC AI500 Benchmark Suite.}
 \begin{threeparttable}
\begin{tabular}[ht]{>{\centering}m{0.11\textwidth}|>{\centering}m{0.11\textwidth}>{\centering}m{0.13\textwidth}>{\centering}m{0.17\textwidth}>{\centering}m{0.13\textwidth}>{\centering}m{0.10\textwidth}>{\centering}m{0.10\textwidth}>{\centering\arraybackslash}m{0.09\textwidth}}
\toprule
\textbf{Problem Domains} & \textbf{Models} & \textbf{Datasets} & \textbf{Target Quality} & \textbf{AI Frameworks} &\textbf{Comm Lib\tnote{1}} &\textbf{AI Acc Lib\tnote{2}} & \textbf{Epochs}  \\
\midrule
EWA & FasterRCNN  & EWA & mAP@[IoU=0.5]\\=0.35 & TensorFlow & Horovod & CUDA, cuDNN, NCCL &50\\ \midrule
Image Classification & ResNet50 v1.5  & ImageNet & TOP 1 Accuracy=0.763 & TensorFlow & Horovod & CUDA, cuDNN, NCCL & 90 \\ 

\bottomrule
\end{tabular}
\begin{tablenotes}
        \footnotesize
        \item[1] Comm Lib refers to the communication libraries. 
        \item[2] AI acc lib refers to AI accelerators libraries. 

      \end{tablenotes}
\end{threeparttable}
\label{table:benchmark_suite}
\vspace{-0.5cm}
\end{table}%

\subsection{The Simple Metric}\label{sec:metrics}

\subsubsection{Valid FLOPS}\label{sec:vflops}
We propose Valid FLOPS (in short, VFLOPS) to quantify the valid performance that considers both the system throughput and model quality. The goal of this metric is to impose a penalty on failing to achieve a target quality.
VFLOPS is calculated according to the formulas as follows.

\begin{equation}
    VFLOPS=FLOPS*penalty\_coefficient \label{formula1}
\end{equation}

The penalty\_coefficient is used to penalize or award the FLOPS if the achieved quality is lower or greater than the target quality. Its definition is described as follows:
\begin{equation}
\begin{split}
    penalty\_coefficient=(achieved\_quality/target\_quality)^{n} \label{formula2}
\end{split}
\end{equation}

Here, $achieved\_quality$ represents the actual model quality achieved in the evaluation. $target\_quality$ is the state-of-the-art model quality that we predefine in our benchmarks~\ref{table:benchmark_suite}. The value of n is a positive integer, which we use to define the model quality's sensitivity. The higher the number of n, the more loss of quality drop. 
EWA has a much more stringent quality requirement than Image Classification; we set n as 10 for EWA and 5 for Image Classification by default. 

Previous work~\cite{tang_ISPASS_2021,mattson2020mlperf} shows most AI tasks are stochastic in terms of the training epochs to achieve a specified target quality. However, for training on a given system, the FLOPS is fixed. According to Equation~\ref{formula1} and Equation~\ref{formula2}, we know that for an AI training workload with fixed epochs, VFLOPS is only related to the achieved quality.  So the variance of the achieved\_quality decides the repeatability of VFLOPS. We have conducted a thorough analysis of the run-to-run variation of AIBench in Sec~\ref{subsec-achieve-repeat} and further select the most repeatable workloads to assure the repeatability of VFLOPS.

\section{Case Study}
This section presents a case study using HPC AI500 benchmark. We perform a series of scaling experiments on a HPC AI system using HPC AI500 benchmark suite to show the scalability of the reference implementation (Sec.~\ref{subsec-scaling-expriments}). We provide an analyze to illustrate why EWA and image classification have different parallel efficiency (Sec.~\ref{subsec-why}). Finally, we publish a VFLOPS ranking list using Image Classification to show the simpleness of this metric (Sec.~\ref{subsec-ranking}). 

\subsection{Experimental Setting}
The experiments are conducted on an HPC AI system, consisting of eight nodes, each of which is equipped with one Intel(R) Xeon(R) Platinum 8268 CPU and eight NVIDIA Tesla V100 GPUs. Each GPU in the same node has 32GB HBM memory, connected by NVIDIA NVLink--a high-speed GPU interconnection whose theoretical peak bi-directional bandwidth is 300GB/s. The nodes are connected with an  Ethernet networking with a bandwidth of 10 Gb/s. Each node has a 1.5 TB system memory and an 8 TB NVMe SSD disk. 

The details of the architecture of each NVIDIA Tesla V100 GPU--NVIDIA Volta architecture are as follows.
The NVIDIA Volta architecture is equipped with 640 Tensor Cores to accelerate GEMM and convolution operations.
Each Tensor Core performs 64 floating-point fused-multiply-add (FMA) operations per clock, delivering up to 125 TFLOPs of theoretical peak performance. When performing mixed precision training with a Tensor Core, we use FP16 for calculation and FP32 for accumulation. 

We use TensorFlow v1.14, compiled with CUDA v10.1 and cuDnn v7.6.2 backend. We use Horovod v0.16.4 for synchronous distributed training, compiled with OpenMPI v3.1.4 and NCCL v2.4.8. 
NCCL is short for the NVIDIA Collective Communications Library. We use NVProf~\cite{nvprof} to measure the FLOPs.

\begin{figure}[ht]
\scriptsize
\begin{subfigure}{0.32\textwidth}
  \centering
  \includegraphics[width=.85\linewidth]{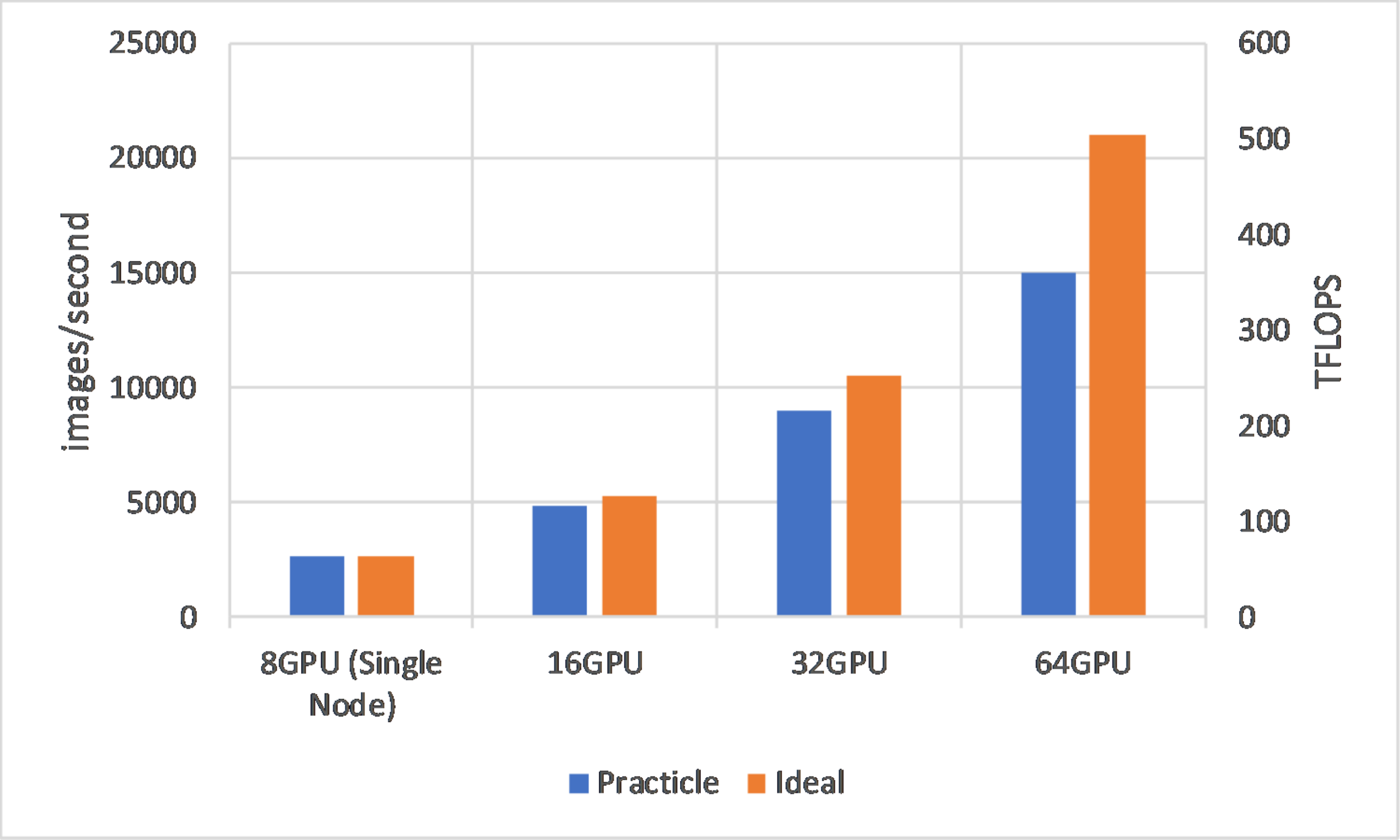}
  \caption{IC (FP32) }
  \label{fig:imagenet-fp32-scaling}
\end{subfigure}
\begin{subfigure}{0.32\textwidth}
  \centering
  \includegraphics[width=.85\linewidth]{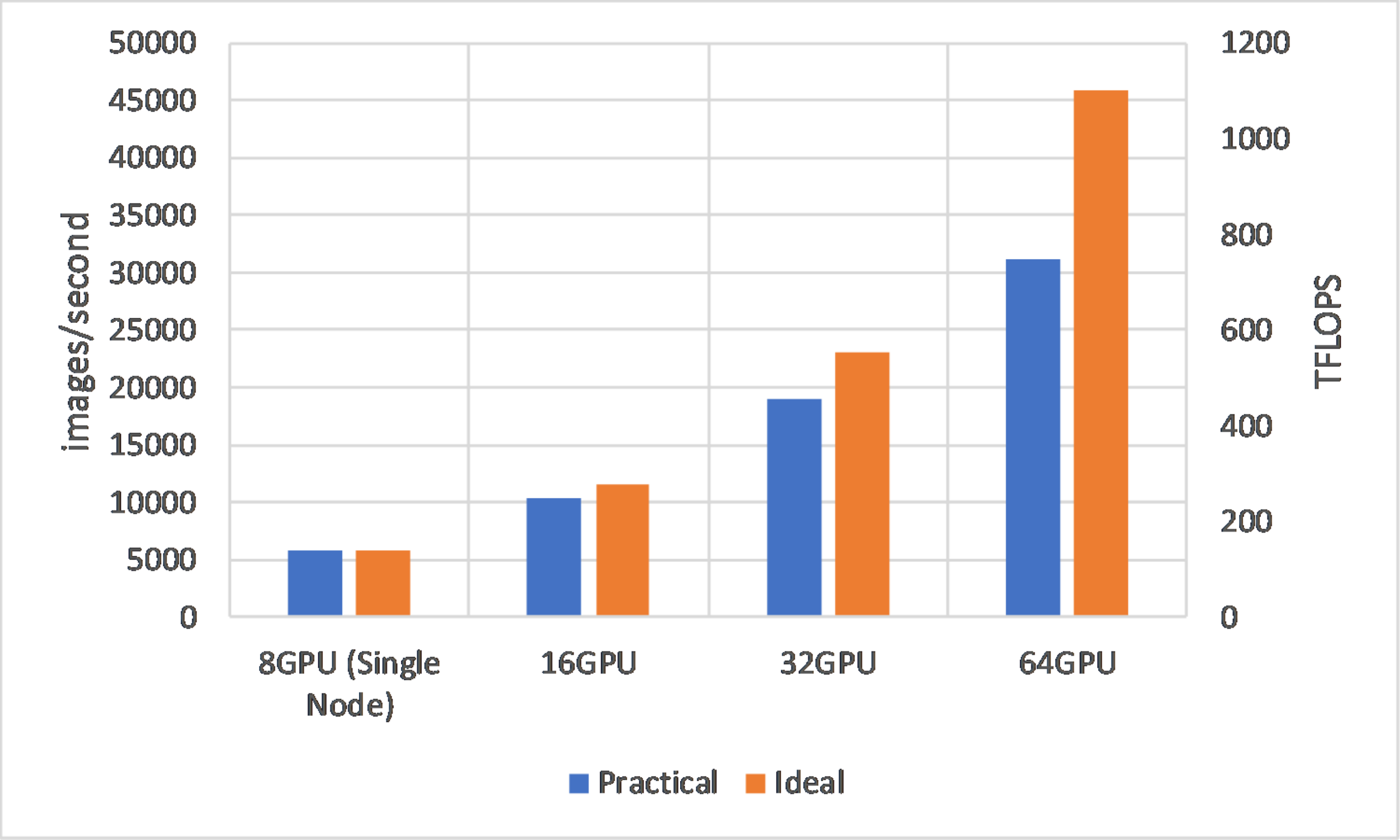}
  \caption{IC (Mixed)}
  \label{fig:imagenet-mixed-scaling}
\end{subfigure}
\begin{subfigure}{0.32\textwidth}
  \centering
  \includegraphics[width=.85\linewidth]{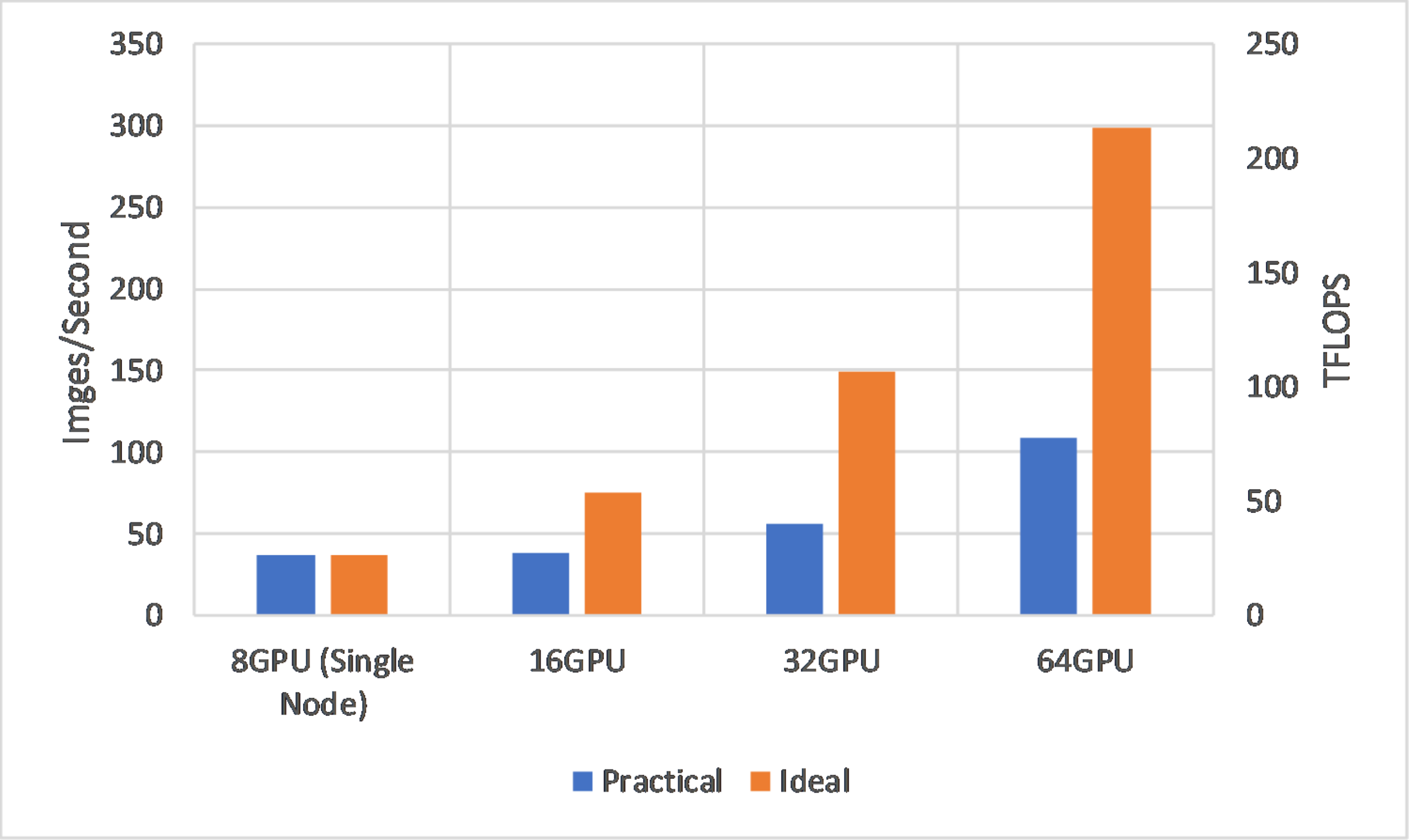}
  \caption{EWA (FP32)}
  \label{fig:EWA-fp32-scaling}
\end{subfigure}

\begin{subfigure}{0.32\textwidth}
  \centering
  \includegraphics[width=.85\linewidth]{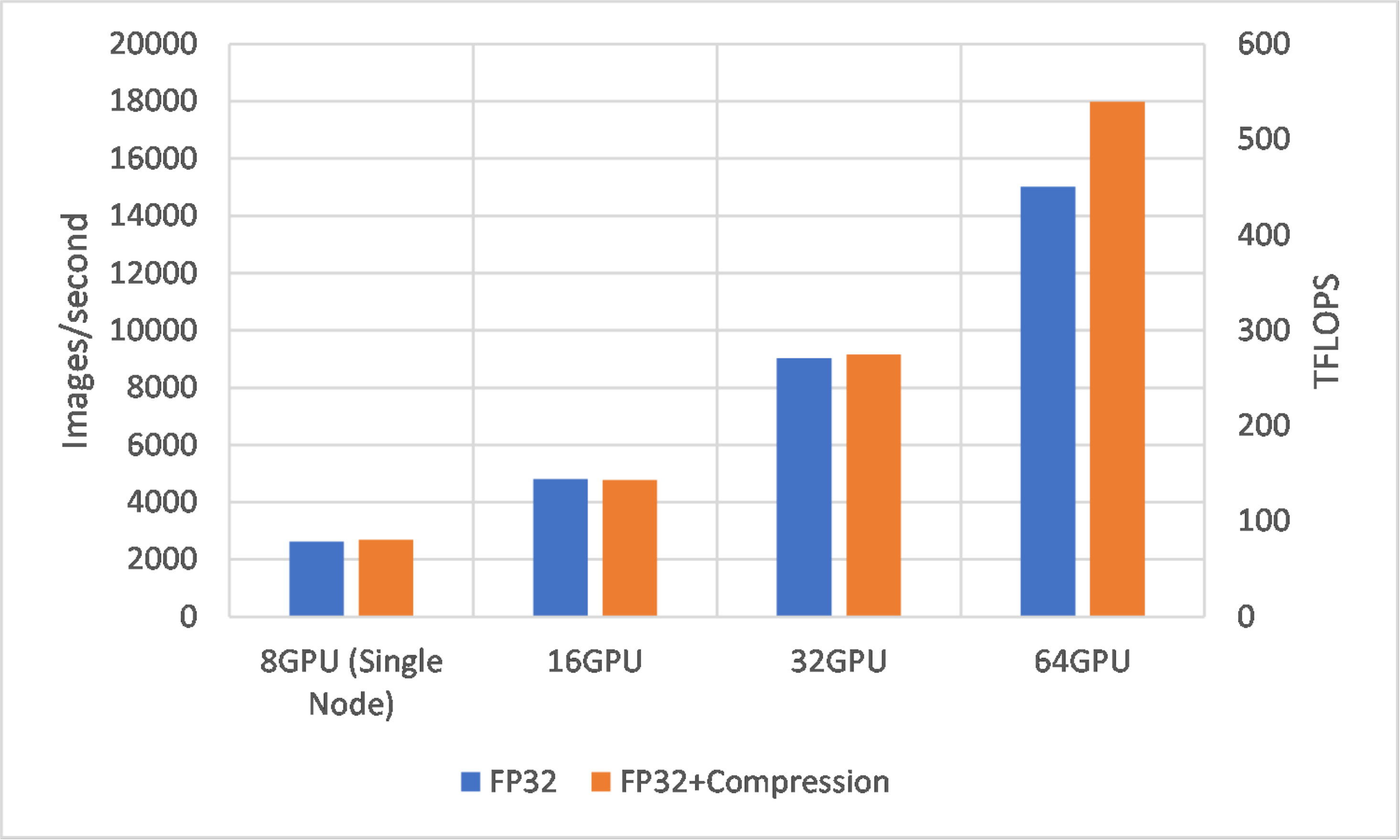}
  \caption{IC (FP32+Compression) }
  \label{fig:imagenet-fp32-compression-scaling}
\end{subfigure}
\begin{subfigure}{0.32\textwidth}
  \centering
  \includegraphics[width=.85\linewidth]{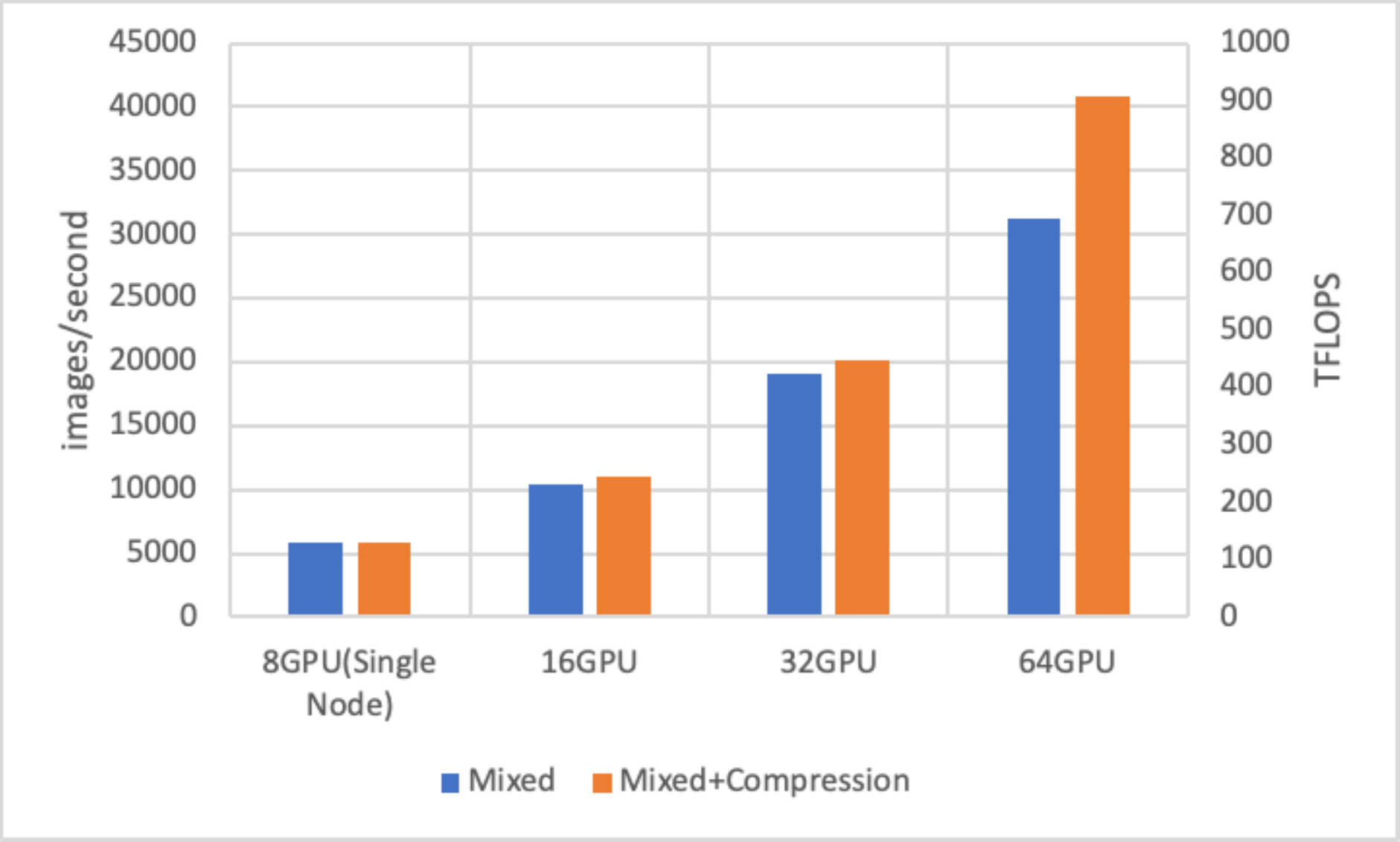}
  \caption{IC (Mixed+Compression)}
  \label{fig:imagenet-mixed-compression-scaling}
\end{subfigure}
\begin{subfigure}{0.32\textwidth}
  \centering
  \includegraphics[width=.85\linewidth]{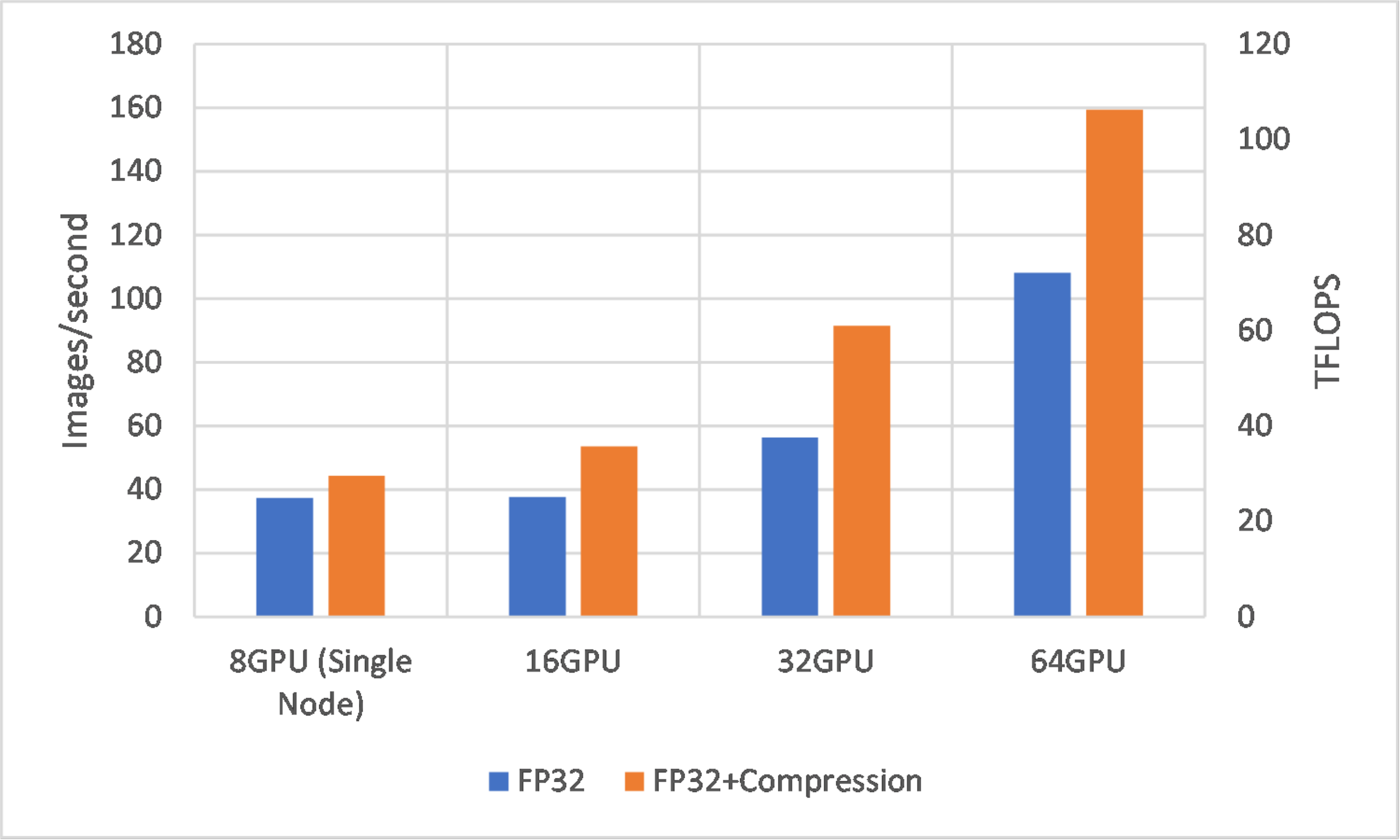}
  \caption{EWA (FP32+Compression)}
  \label{fig:EWA-fp32-compression-scaling}
\end{subfigure}
\caption{The scaling experiments of Extreme weather analysis (EWA) and Image 
Classification (IC).  
}
\vspace{-0.5cm}
\label{fig:scaling}
\end{figure}

\subsection{Scaling Experiments of HPC AI500 Benchmarks}~\label{subsec-scaling-expriments}
Both EWA and Image Classification experiments are scaled out from 8 GPUs to 64 GPUs. In addition to the original FP32 version, we also evaluate performance of  the mixed-precision and the model compression version, which are two frequently-used optimizations in HPC AI applications~\cite{kurth2018exascale, jia2018highly}. We take the 8-GPU experiments (single node) as a baseline. Our communication topology is the double binary tree, which is implemented by NCCL 2.4.  We report the performance numbers of these experiments and perform further analysis. The scaling results are shown in Fig.~\ref{fig:scaling}.

\subsubsection{Image Classification} 
 For the FP32 precision implementation of Image Classification, the parallel efficiency is 0.91, 0.85, and 0.71 on 16, 32, and 64 GPUs, respectively. For the mixed implementation, the parallel efficiency is slightly lower: 0.89, 0.82, and 0.67, respectively. There is a significant loss of parallel efficiency when the system scale is 64 GPUs. The reason is that when the system scales up to 64 GPUs (8 nodes), more data need to be transmitted over the low-speed Ethernet and result in the parallel efficiency reduction. 
 
 We also notice that communication compression does not improve the performance when the system scale is 32 GPUs or less. This is because that communication is not the bottleneck under these situations. However, when the scale is 64 GPUs, it contributes a lot. For the FP32 version, the performance improves from 345 to 414 TFLOPS. For the mixed version, the performance improves from 718 to 939 TFLOPS.  
 
 The highest performance of Image Classification that we achieve is 939 TFLOPS through both mixed precision optimization and communication compression, as shown in Fig.~\ref{fig:imagenet-mixed-compression-scaling}.    
 
\subsubsection{EWA}
For the FP32 precision implementation of EWA, the parallel efficiency is 0.50, 0.37, and 0.36 at the system scale of 16, 32, and 64 GPUs, respectively. As a communication-intensive workload, communication compression of EWA achieves good results. When communication compression is used, the performance gain persists when the scale increase from 8 to 16, 32, and 64 GPUs, and the speedup is 1.2, 1.4,1.6 and 1.5, respectively. The highest performance of EWA achieved through communication compression is 109 TFLOPS.

\subsection{Why EWA and Image Classification Have Different Parallel Efficiencies?} \label{subsec-why}

\begin{figure}[hbt!]
    \centering
    \includegraphics[width=.6\linewidth]{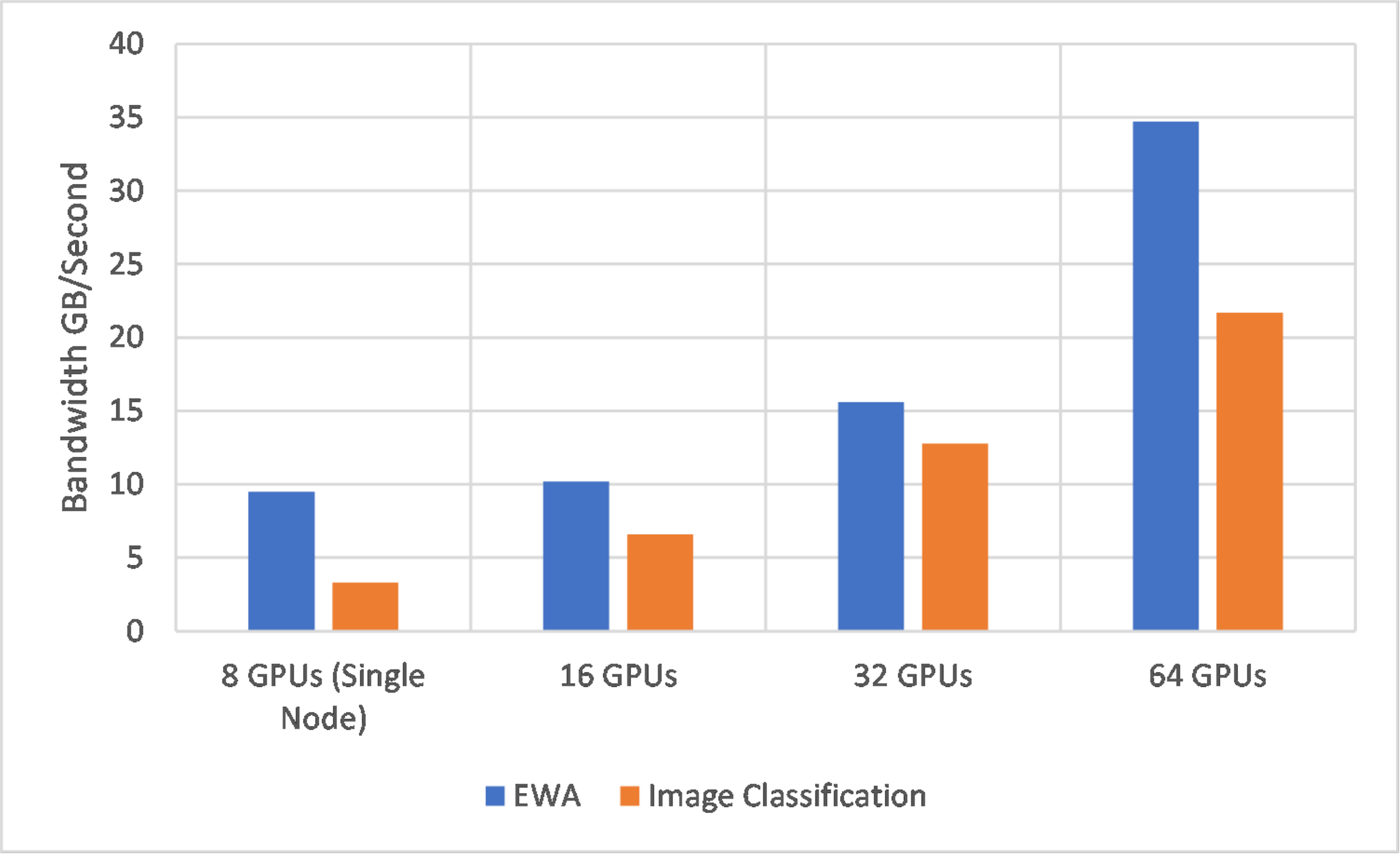}
    \caption{The distinctive communication bandwidth consumption of the FP32 implementations of EWA and Image Classification. The bandwidth refers to the aggregated communication bandwidth in the systems, including Ethernet communication between nodes and NVLink communication between GPUs within the node.}
    \label{fig:bandwidth-ewa-imagenet}
    \vspace{-0.5cm}
\end{figure}

We found their different parallel efficiencies for EWA and Image Classification due to distinct communication bandwidth consumption. As shown in Fig.~\ref{fig:bandwidth-ewa-imagenet}, we measure the communication bandwidth consumption of the FP32 precision implementations of EWA and Image Classification. EWA consumes a much higher communication bandwidth than that of Image Classification. It shows that EWA is a communication-intensive workload. In contrast, the Image Classification is a computation-intensive workload. 

\subsection{VFLOPS Ranking of HPC AI500 Systems using Image Classification}\label{subsec-ranking}
The metric of VFLOPS emphasizes both performance and quality. 
We have published the VFLOPS ranking of Image Classification in HPC AI500 ranking website:~\url{http://www.benchcouncil.org/HPCAI500/ranking.html}. Fujitsu won first place by achieving 31.41 VPFLOPS. Meanwhile, another additional metric--time-to-quality is also reported. Generally, our metric is visual and straightforward.

\section{Related Work}\label{related_work}

We review the recent efforts of HPC AI benchmarking in chronological order.

HPC AI500 (V 1.0) (2018)~\cite{jiang2018hpc} is the first HPC AI benchmarks based on the real-world scientific dataset, covering high energy physics, cosmology, and extreme weather analytics. HPC AI500 directly extracts three benchmarking scenarios from representative HPC AI applications without a systematic benchmarking methodology to produce a representative, repeatable and simple benchmark suite.

 HPL-AI (2019)~\cite{hplai} is designed for 32-bit and even lower floating-point precision AI computing. It uses LU decomposition of mixed precision as the core algorithm. As a micro benchmark, HPL-AI enables repeatable evaluation and easily be ported to different systems. However, it cannot provide model-level evaluation using an entire training session and lacks relevance in the AI field. Just like we have discussed in Sec.~\ref{challenge}, the benchmark results will be misleading.

Deep500 (2019)~\cite{deep500} is a reproducible customized benchmarking infrastructure for high-performance deep learning. It has four abstraction levels to provide a full-stack evaluation and provides a simple reference implementation based on small data sets and models. Deep500 can customize workloads but lacks benchmark specifications. It is more like a framework than a concrete benchmark.

\section{Conclusion}

This paper proposes a representative, repeatable, and simple HPC AI benchmarking methodology. We analyze the representativeness and repeatability of AIBench. After further considering the HPC field's additional requirements, we build the HPC AI500 benchmark suite, containing the two most representative, repeatable workloads, namely extreme weather analysis (EWA) and image classification. We propose 
Valid FLOPS
to rank the performance 
of HPC AI systems. We evaluate an HPC AI system using the HPC AI500 benchmarks, showing the reference implementation's scalability. Also, we publish a VFLOPS-based ranking list. The specification, source code, and HPC AI500 ranking numbers are publicly available from \url{http://www.benchcouncil.org/benchhub/hpc-ai500-benchmark}. A full technical report is available from ~\cite{jiang2020hpc}.

\bibliographystyle{ieeetr}

\end{document}